%% file: paper.tex
\documentclass[runningheads,a4paper]{sig-alternate-10pt}
\usepackage{graphicx}
\usepackage{caption}
\usepackage{subfigure}
\usepackage{paralist}
\usepackage{multirow}
\usepackage{url}
\usepackage{rotating}
\usepackage{float}
\usepackage{array}
\usepackage{mathtools}
\usepackage{algorithm}
\usepackage{algorithmic}

\begin{document}



\title{Fighting Authorship Linkability with Crowdsourcing}
\author{
\alignauthor 
Mishari Almishari$^\ddag$,  Ekin Oguz$^\dag$, Gene Tsudik$^\dag$\\ 
\affaddr{$^\ddag$ King Saud University\\$^\dag$University of California, Irvine}
}

\maketitle

\begin{abstract}
Massive amounts of contributed content -- including traditional literature, blogs, music, videos, reviews and tweets --  
are available on the Internet today, with authors numbering in many millions. Textual information, such as
product or service reviews, is an important and increasingly popular type of content that is being
used as a foundation of many trendy community-based reviewing sites, such as TripAdvisor and Yelp.
Some recent results have shown that, due partly to their specialized/topical nature, sets of reviews
authored by the same person are readily linkable based on simple stylometric features.
In practice, this means that individuals who author more than a few reviews under different accounts (whether
within one site or across multiple sites) can be linked, which represents a significant loss of privacy.

In this paper, we start by showing that the problem is actually worse than previously believed.
We then explore ways to mitigate authorship linkability in community-based reviewing.
We first attempt to harness the global power of crowdsourcing
by engaging random strangers into the process of re-writing reviews. As our empirical
results (obtained from Amazon Mechanical Turk) clearly demonstrate, crowdsourcing yields impressively
sensible reviews that reflect sufficiently different stylometric characteristics such
that prior stylometric linkability techniques become largely ineffective. We also consider using
machine translation to automatically re-write reviews. Contrary to what was previously believed, 
our results show that translation decreases authorship linkability as the number of intermediate languages grows.
Finally, we explore the combination of crowdsourcing and machine translation and report on the results.
%
%
\end{abstract}

\input{intro.tex}

\input{related.tex}
\input{background.tex}
\input{dataset-setting.tex}
\input{experiments.tex}
\input{crowdsourcing.tex}
\input{translation.tex}
\input{discussion}
\input{conclusion.tex}

\newpage
\bibliographystyle{abbrv}
\bibliography{paper}

\newpage
\appendix

\input{appendix.tex}

\end{document}

%% file: intro.tex
\section{Introduction}
\label{sec:intro}
The Internet has become a tremendous world-wide bazaar where massive amounts of information
(much of it of dubious quality and value) are being disseminated and consumed on a constant basis.
Sharing of multimedia content is one of the major contributors to Internet's growth and popularity. 
Another prominent  source of shared information is textual, e.g., blogs, tweets and various discussion fora.
Among those, community reviewing has carved out an important niche. This category includes
well-known sites, such as: Yelp, CitySearch, UrbanSpoon, Google Places and TripAdvisor.
There are also many others that include customer-based reviewing as a side-bar, e.g., Amazon or Ebay.

Regardless of their primary mission and subject coverage, community reviewing sites are popular
since many are free and contain lots of useful content voluntarily contributed by regular people who
document their experience with products, services, destinations, and attractions.
Larger sites, e.g., TripAdvisor and Yelp, have tens of millions of users (readers) and millions of 
contributors~\cite{yelpstat}.

Certain features distinguish community reviewing sites from other contributory Internet 
services:
\begin{itemize}
\item Discussion Fora: these vary from product or topic discussions to comment sections in 
on-line news media. They are often short and not very informative (even hostile). 
\item Body of Knowledge: the best-known and most popular example is Wikipedia -- a huge amalgamation 
of communal knowledge on a very wide range of subjects. However, unlike reviewing sites where
each review is atomic and discernable, related contributions to body-of-knowledge sites 
are usually mashed together, thus (by design) obscuring individual prose. 
\item Online Social Networks (OSNs): such sites are essentially free-for-all as far as the type 
and the amount of contributed information. Since most OSNs restrict access to content provided by a user to 
``friends'' (or ``colleagues'') of that user, opinions and reviews do not propagate 
to the rest of Internet users.
\end{itemize}
Some recent work \cite{mish-linkability} has shown that many contributors
to community reviewing sites accumulate a body of authored content that is sufficient for creating
their stylometric 
profiles, based on rather simple features (e.g., digram frequency). A stylometric profile allows
probabilistic linkage among reviews generated by the same person.
This could be used to link reviews from different accounts (within a site or across sites) operated by the same user. 
On one hand, tracking authors of spam reviews can be viewed as a useful service. On the other hand, the ease of 
highly accurate linkage between different accounts is disconcerting and ultimately detrimental to
privacy. Consider, for example, a vindictive merchant who, offended
by reviews emanating from one account, attempts to link it to other accounts held by the same person,
e.g., for the purpose of harrassment. 
We consider both sides of this debate to be equally valid and do not choose sides. 
However, we believe that the privacy argument deserves to be considered,
which triggers the motivation for this paper: 
\vspace{2 mm}
\\
\centerline{\fbox{\begin{minipage}{0.46\textwidth}
\centering
\it
What can be done to mitigate linkability of reviews authored by
the same contributor?
\end{minipage}}}
\\
\paragraph{Roadmap:} 
Our goal is to develop techniques that mitigate review account linkability. 
To assess efficacy of proposed techniques, we need accurate 
review linkage models. To this end, we first improve state-of-art author review linkage methods. 
We construct a specific technique, that offers $90\%$ accuracy,  even for a small 
number of identified reviews (e.g., $95$) and a smaller set (e.g., $5$) of anonymous reviews.

Our second direction is the exploration of techniques that decrease authorship linkability. 
We start by considering crowdsourcing, which entails engaging random strangers in rewriting
reviews. As it turns out, our experiments using Amazon MTurk \cite{amazonmturk} clearly 
demonstrate that  authorship linkability can be significantly inhibited by crowdsourced rewriting. 
Meanwhile, somewhat surprisingly, crowd-rewritten reviews  remain meaningful and generally 
faithful to the originals. We then focus on machine translation tools and show that,  by randomly 
selecting languages to (and from) which to translate, we can substantially decrease linkability.
\paragraph{Organization:} 
The next section summarizes related work.
Then, Section \ref{sec:background} overviews some preliminaries, followed by
Section \ref{sec:dataset-settings} which describes the experimental dataset and review 
selection process for subsequent experiments. 
Next, Section \ref{sec:link-analysis} discusses our linkability study and its outcomes.
The centerpiece of the paper is Section \ref{sec:red-auth-link}, which presents  
crowdsourcing and translation experiments. It is followed by Section \ref{sec:discussion}
where we discuss possible questions associated with the use of crowdsourcing.
Finally, summary and future work appear in Section \ref{sec:conclusion}.

%% file: related.tex
\hspace{1cm}
\section{Related Work}
\label{sec:related}
Related work generally falls into two categories: Authorship Attribution/Identification and
Author Anonymization.

{\bf Authorship Attribution.} There are many studies in the literature. For example, \cite{mish-linkability} 
shows that many Yelp's reviewers are linkable using only very simple feature set.
While the setting is similar to ours, there are some notable differences. First, we obtain high linkability 
using very few reviews per author. Second, we only rely on features extracted 
from review text. A study of blog posts achieves 80\% linkability accuracy \cite{internet-scale}.  
Author identification is also studied in the context of academic paper reviews achieving accuracy of 90\% \cite{herbert-deanonymizer}. 
One major difference between these studies and our work
is that we use reviews, which are shorter, less formal and less restrictive in choice of words than 
blogs and academic papers. Abbasi and Chen propose a well-known author attribution technique based on 
Karhunen-Loeve transforms to extract a large list of Writeprint features (assessed in 
Section \ref{sec:link-analysis}) \cite{writeprints-abbasi}. Lastly, Stamatatos 
 provides a comprehensive overview of authorship attribution studies \cite{author-survey}.

{\bf Author Anonymization.} There are several well-known studies in author anonymization \cite{pseudonymity,obfuscate-stylo,anonymouth}.  
Rao and Rohatgi are among the first to address authorship anonymity by proposing using round-trip machine 
translation, e.g., English $\rightarrow$ Spanish $\rightarrow$ English, to obfuscate authors \cite{pseudonymity}. 
Other researchers apply round-trip translation, with
a maximum of two intermediate languages and show that it does not provide noticeable 
anonymizing effect \cite{translate-once-twice,brennan2012adversarial}. In contrast, we explore effects (on privacy) of increasing and/or randomizing 
the number of intermediate languages. 

Kacmarcik and Gamon show how to anonymize documents via obfuscating writing style,
by proposing adjustment to document features to reduce the 
effectiveness of authorship attribution tools \cite{obfuscate-stylo}. The main limitation 
of this technique is that it is only applicable to authors with a fairly large
text corpus, whereas, our approach is applicable to authors with limited number of reviews.

Other practical-counter-measures for authorship recognition 
techniques such as obfuscation and imitation
attacks are explored \cite{attack-against-recognition}. However, it is shown that  such stylistic deception can be detected 
with $96.6\%$ accuracy \cite{detect-fraud}.

The most recent relevant work is Anonymouth \cite{anonymouth} --
a framework that captures the most effective features of documents for linkability and 
identifies how these feature values should be changed to achieve anonymization. 
Our main advantage over Anonymouth is usability. Anonymouth requires the author to have two 
additional sets of  documents, on top of the original document to be anonymized: 1) sample 
documents written by the same author and 2) a corpus of sample documents written by 
other authors. Whereas, our approach does not require any such sets. 

%% file: background.tex
\section{Background}
\label{sec:background}
This section overviews stylometry, stylometric characteristics and statistical 
techniques used in our study. 

Merriam-Webster dictionary defines \texttt{Stylometry} as:
{\em the study of the chronology and development of an author's work based 
especially on the recurrence of particular turns of expression or trends of 
thought.} We use stylometry in conjunction with the following two tools:
\begin{compactitem}
  \item [\bf Writeprints feature set:] well-known stylometric features used to
  analyze author's writing style. 
  \item [\bf Chi-Squared test:] a technique that computes the distance between
  each author's review in order to assess linkability.
\end{compactitem}

\subsection{Writeprints}
\label{wp-background}
Writeprints is essentially a combination of static and dynamic stylometric features that 
capture lexical, syntactic, structural, content and idiosyncratic properties of a 
given body of text \cite{writeprints-abbasi}. Some features include:
\begin{itemize}
  \item Average Character Per Word: Total number of characters divided by total number of words.
  \item Top Letter Trigrams: Frequency of contiguous sequence of $3$ characters,
  e.g. $aaa, aab, aac, ..., zzy,$ $zzz$. There are $17576$ ($26^3$) possible 
  permutation of letter trigrams in English.
  \item Part of Speech (POS) Tag Bigrams: POS tags are the mapping of words to
  their syntactic behaviour within sentence, e.g. noun or verb. POS tag bigrams
  denotes $2$ consecutive parts of speech tags. We used Stanford POS Maxent 
  Tagger~\cite{stanfordpos} to label each word with one of $45$ possible POS tags.
  \item Function Words: Set of $512$ common words, e.g. $again$, $could$, $himself$
  and etc, used by Koppel et al. in Koppel, 2005.
\end{itemize}

Writeprints has been used  
in several stylometric studies \cite{writeprints-abbasi,internet-scale,herbert-deanonymizer}.
It has been shown to be an effective means for identifying authors because
of its capability to capture even smallest nuances in writing.

We use Writeprints implementation from JStylo -- a Java library that includes
$22$ stylometric features \cite{anonymouth}.

\subsection{Chi-Squared Test}
\label{cs-background}
Chi-Squared (CS) test is used to measure the distance between two distributions \cite{chi-square}. 
For any two distributions $P$ and $Q$, it is defined as:
\begin{equation}
CS_{d}(P,Q)=\sum_i\frac{(P(i)-Q(i))^2}{P(i)+Q(i)}\nonumber
\label{eq:cs}
\end{equation}
$CS_d$ is a symmetric measure, i.e., $CS_d(P,Q)$ = $CS_d(Q,P)$. Also, it is always non-negative;
a value of zero denotes that $P$ and $Q$ are identical distributions. We employ Chi-Squared 
test to compute the distance between contributor's anonymous and identified reviews.

\if{0}
\subsection{Jaccard Index}
\label{ji-background}
Jaccard index (also knowns as the similarity coefficient) is a widely used measure for capturing 
similarity and diversity of sets \cite{jaccard}. For any two sets $A$ and $B$, Jaccard index 
is defined as:
$$
J(A,B)=\frac{|A \cap B |}{|A \cup B|} 
\label{eq:js}
$$
Its value ranges between zero and one, where zero denotes complete divergence and 
one denotes complete similarity. Jaccard distance, which measures diversity of two 
sets, is defined as: 
\begin{equation}
J_\delta(A,B)=1 - \frac{|A \cap B |}{|A \cup B|} = 1-J(A,B) 
\label{eq:jd}
\end{equation}
$J(A, B)$ and $J_\delta(A, B)$ are useful for evaluating 
similarity and diversity of original and rewritten reviews.
\fi

%% file: dataset-setting.tex
\section{Linkability Study Parameters}
\label{sec:dataset-settings}
This section describes the dataset and problem setting for our linkability analysis.

\subsection{Dataset}
\label{sec:dataset}
We use a large dataset of reviews from Yelp\footnote{See: \url{www.yelp.com}.} with
$1,076,850$ reviews authored by $1,997$ distinct contributors.
We select this particular dataset for two reasons:

\begin{enumerate}
  \item Large number of authors with widely varying numbers of reviews: average 
  number of reviews per author is $539$, with a standard deviation of $354$.
  \item Relatively small average review size -- $149$ words -- which should make 
  linkability analysis more challenging.
\end{enumerate}

\subsection{Problem Setting}
For a given anonymous set of reviews $R$, we want to link them to a set of identified reviews -- with a known author. The problem becomes challenging when the set of anonymous and identified reviews are relatively small. The exact problem setting is as follows:

We first select $40$ authors at random. We pick this relatively small number 
in order to make subsequent crowdsourcing experiments feasible, as described
in Section \ref{sec:red-auth-link}. Then, for each author, we randomly shuffle her reviews and select the first $N$ ones. Next, we split the selected reviews into two sets:

\begin{itemize}
  \item First $X$ reviews form the \textbf{Anonymous Record} (AR) set. 
  We experiment with AR sets of varying sizes.
  \item Subsequent $(N-X)$ reviews form the \textbf{Identified Record} (IR) set.
\end{itemize}

Our problem is reduced to linking ARs to their corresponding IRs.
We set $N$=$100$ and we vary $X$ from $1$ to $5$.
This makes our IRs and ARs quite small compared to an average of $539$ reviews per author in the dataset.
As a result, our linking problem becomes very challenging.

Next, we attempt to link ARs to their corresponding IRs.
More specifically, for each AR, we rank -- in descending order of likelihood -- all possible authors (i.e. IRs).
Then, the top-ranked IR (author) represents the most similar IR to the given AR.
If the correct author is among top-ranked $T$ IRs, linking model has a hit; otherwise, it has a miss.
For a given value of $T$, we refer to the fraction of hits of all ARs (over the total of $40$) as Top-$T$ linkability ratio.
Our linkability analysis boils down to finding a model that maximizes this linkability ratio for different $T$ and $AR$ sizes. We consider two integer values of $T$: $[1;4]$, where $1$ denotes a perfect-hit and $4$ stands for an almost-hit.

%% file: experiments.tex
\section{Linkability Analysis}
\label{sec:link-analysis}
We first apply a subset of the popular Writeprints feature set\footnote{We initially experimented
with the Basic-9 feature set, which is known to  provide useful information for author identification
for less than $10$ potential authors \cite{brennan2012adversarial}.
However, its performance was really poor, since we have $40$ authors in our smallest set.}
to convert each AR and IR into a token set. We then use Chi-Square\footnote{We tried others tests
including: Cosine, Euclidean, Manhattan, and Kullback-Leibler Divergence. 
However, Chi-Squared Test outperformed them all.} to compute distances between those token sets. 
We now describe our methodology in more detail. Notation and abbreviations are reflected in Table \ref{tab:notation}. 
\begin{table}[t!]
\centering
\begin{tabular}{|r|p{0.31\textwidth}|}
\hline
LR & Linkability Ratio \\ \hline
AR & Anonymous Records \\ \hline
IR & Identified Records \\ \hline
CS & Chi-Squared Distance Model \\  \hline
$F$ & A feature \\ \hline
$F_T$ & The set of tokens in feature $F$ \\ \hline
$S_F$  & Set of selected features \\ \hline
$WP_i$  & Writeprint feature i \\ \hline
$WP_{all}$ & Combination of all Writeprints \\ \hline
$CS_d(IR,AR)$  & CS distance between $IR$ and $AR$ \\ \hline
\end{tabular}
\caption{Notation and abbreviations}
\label{tab:notation}
\end{table}

\subsection{Methodology}
\label{method}
Firstly, we tokenize each AR and IR sets using every feature -- $F$ -- in our set of selected features -- $S_F$ -- to obtain a 
set of tokens $F_T=\{F_{T_1}, F_{T_2}, ... , F_{T_n}\}$, where  $F_{T_i}$ denotes the 
$i$-th  token in $F_T$. Then, we compute distributions for all tokens. Next, we use 
CS model to compute the distance between AR and 
IR using respective token distributions. Specifically, to link AR with respect to some feature $F$, 
we compute $CS_d$ between the distribution of tokens in $F_T$ for AR and the distribution 
of tokens in $F_T$ for each IR. After that, we sort the distances in ascending order 
of $CS_d(IR,AR)$ values and return the resulting list. First entry corresponds to the 
IR with the closest distance to AR, i.e., the most likely match.
For more generality in our analysis, we repeat this experiment
$3$ times, randomly picking different AR and IR sets each time. Then, we average out the results.
Note that $S_F$ is initially empty and we gradually add features to it, as described next. 


\begin{table}[t]\small
\centering
\begin{tabular}{cc|c|c|c|} 
\cline{3-4}
& & \multicolumn{2}{ c|}{\textbf{Linkability Ratio}}  \\ \hline
\multicolumn{1}{ |c| } {\textbf{Ranking}} & \textbf{Feature} & \textbf{Top-1(\%)} & \textbf{Top-4(\%)} \\ \hline
\multicolumn{1}{ |c| } {1} & Top Letter Trigrams & 91 & 96  \\ \hline
\multicolumn{1}{ |c| } {2} & POS Bigrams & 89 & 96 \\ \hline
\multicolumn{1}{ |c| } {3} & Top Letter Bigrams & 86 & 94  \\ \hline
\multicolumn{1}{ |c| } {4} & Words & 79 & 94 \\ \hline
\multicolumn{1}{ |c| } {5} & POS Tags & 78 & 90 \\ \hline
\multicolumn{1}{ |c| } {9} & $WP_{all}$ & 52.5 & 82.5 \\ \hline
\end{tabular}
\caption{LRs of best five Writeprint features individually and $WP_{all}$, with $|AR|=5$}
\label{tab:baseline}
\end{table}

\begin{figure*}[t]
  \centering
  \subfigure[Top-1]{\label{fig:training_set_add_pos_top1-png}\includegraphics[scale=0.47]{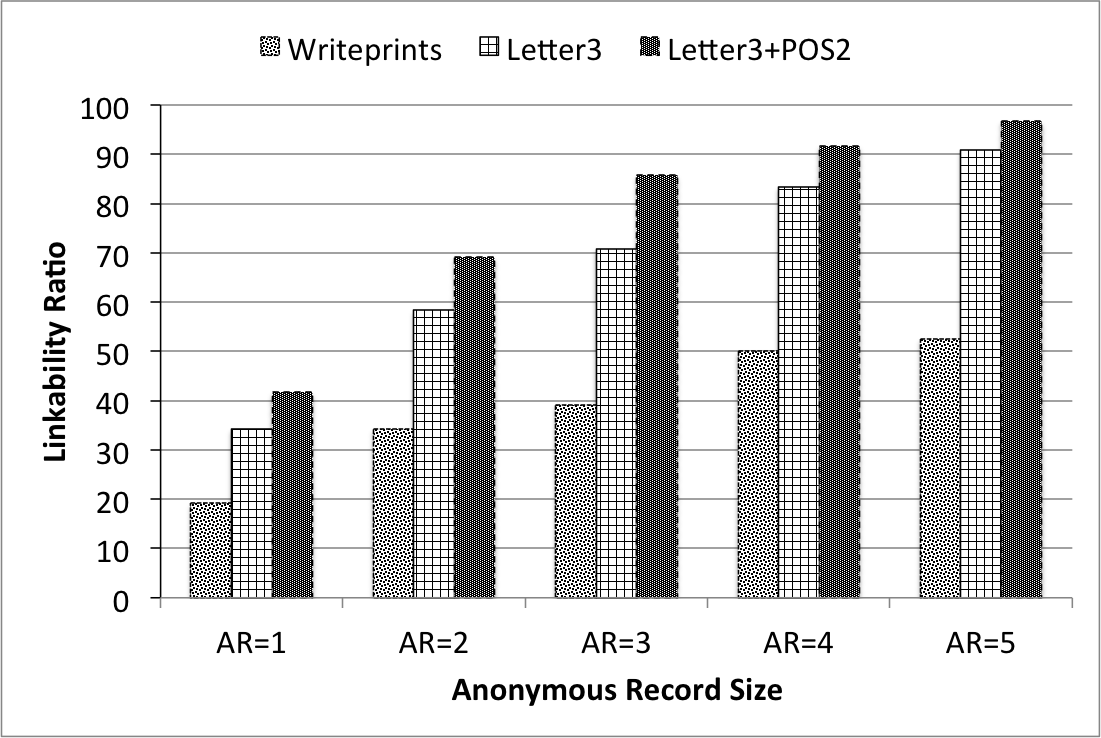}}
  \subfigure[Top-4]{\label{fig:training_set_add_pos_top4-png}\includegraphics[scale=0.47]{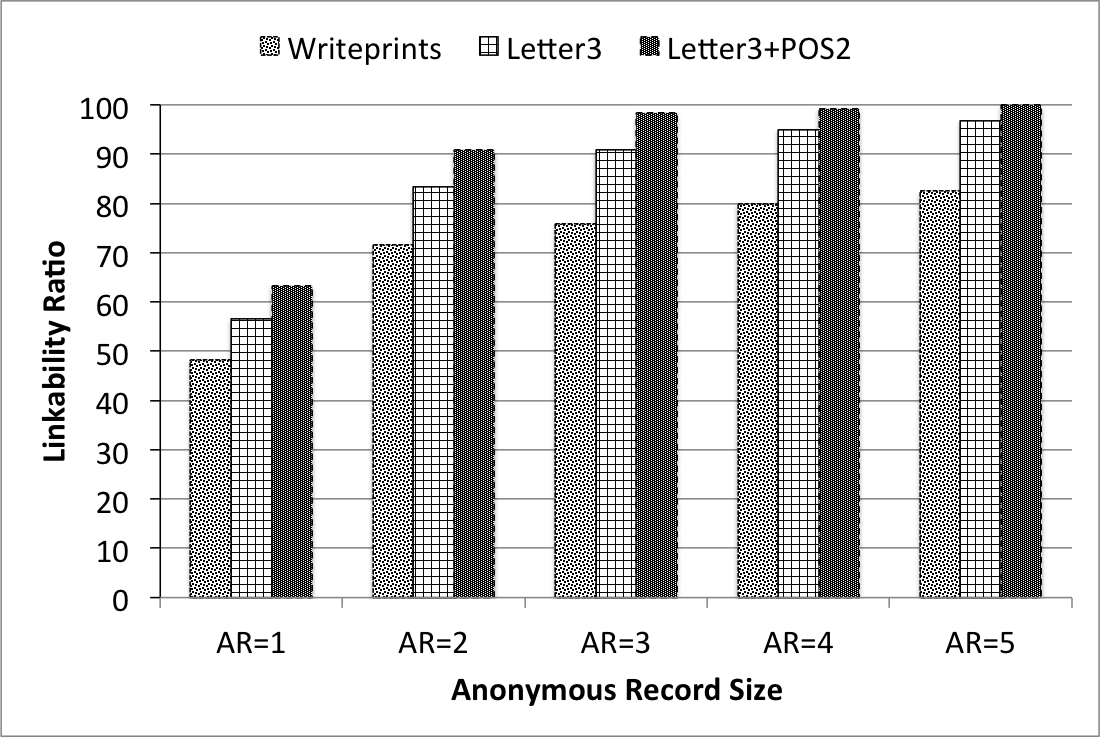}}
  \subfigure[LRs of Letter3+POS2 in varying size of author sets]
    {\label{fig:training-authors}\includegraphics[scale=0.43]{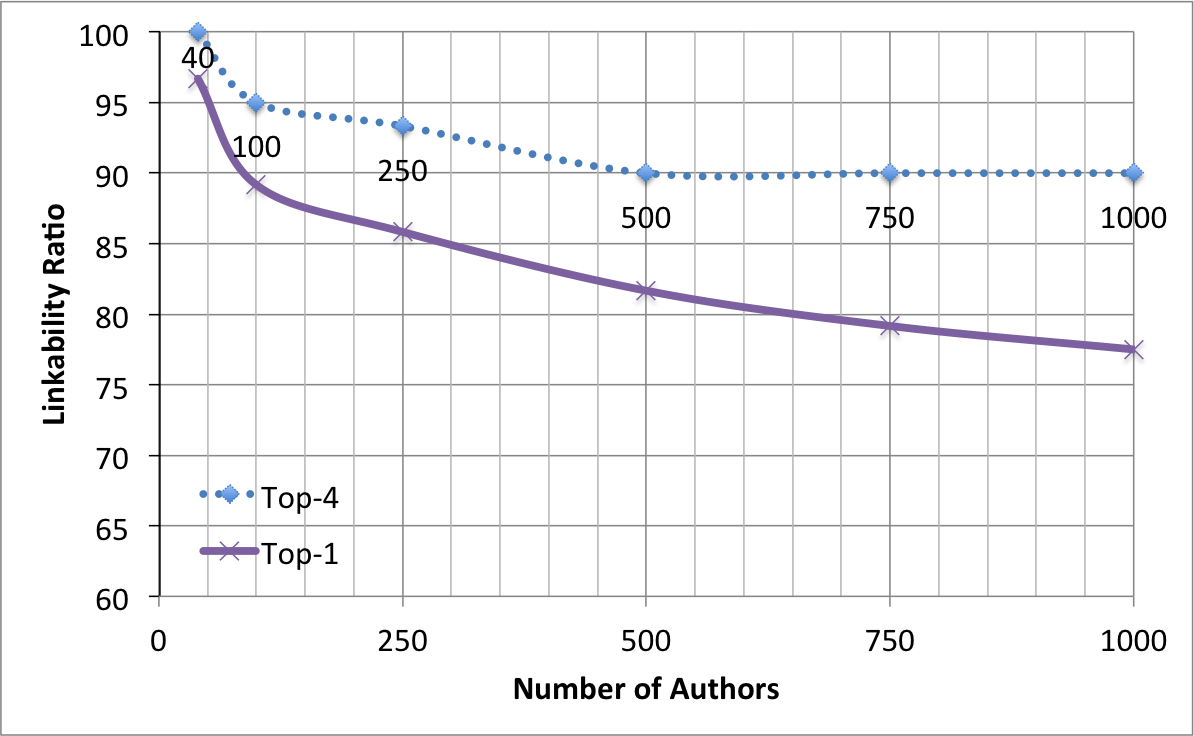}}
  \caption{LRs of Writeprints, Letter3 and Letter3+POS2}
  \label{fig:training-add-pos}
\end{figure*}

\subsection{Feature Selection}
\label{sec:feature-selection}
We use a general heuristics, a version of greedy hill-climbing algorithm, for feature selection \cite{pudil1994floating}.
The idea is to identify most influential features and gradually combine them in $S_F$,
until we encounter a high LR.

\subsubsection{$WP_{all}$}
\label{itm:fs:1st}
As a benchmark, we start with setting feature set $S_F$ to $WP_{all}$,
which combines all $22$ Writeprint features. We compute LR using $WP_{all}$ in CS model
with $|AR|=5$. Unfortunately, $WP_{all}$ results in low LRs -- only $52.5\%$ in Top-1 
and $82.5\%$ in Top-4. We believe that, because of small AR set, combination of 
many features increases noise, which, in turn, lowers linkability.

\subsubsection{Improving $WP_{all}$}
\label{itm:fs:2nd}
Next, we use each feature from $WP_{all}$ individually. That is, we try each $WP_i$ (for
$1\leq~i\leq~22$) with $|AR|=5$. Table \ref{tab:baseline} shows  
the best five features together with $WP_{all}$ after ranking LRs
in Top-1 and Top-4. First five features perform significantly better
than all others, especially, better than $WP_{all}$ which wound up in
$9$-th place. Interestingly, LR increases drastically -- from $52.5\%$ to $91\%$ in Top-1 --
with the best feature. Since Top Letter Trigrams performs best individually, we add it to $S_F$.
Then we move on to considering combination of other four features with Top Letter Trigrams.

\subsubsection{Improving Top Letter Trigrams}
\label{itm:fs:3rd}
Next, we try combining each feature from the set \{POS Bigrams, Top Letter Bigrams, Words, POS Tags\} 
with $S_F$ to check whether it produces a higher LR. 
It turns out that combining POS Bigrams yields the best LR gain: from $91\%$ to $96\%$ in Top-1, and $96\%$ to $100\%$ 
in Top-4. Since we achieve $100\%$ LR in Top-4, we set $S_F$ as \{Top Letter Trigrams, POS Bigrams\}.

We present LR comparisons of experimented features with varying AR sizes in 
Figure \ref{fig:training_set_add_pos_top1-png} and Figure \ref{fig:training_set_add_pos_top4-png}.
For all AR sizes, we notice a significant improvement with Top Letter Trigrams over Writeprints, and
similarly with \{Top Letter Trigrams, POS Bigrams\} over only Top Letter Trigrams in both Top-1 and Top-4.

\subsection{Scalability of the Linkability Technique}
\label{sec:multiauthors}
So far, we assessed linkability of $40$ authors only in a set of $40$ possible authors.
This is partly because computation of $WP_{all}$ in bigger author sets is very expensive.
However, $40$ is a really small size in a real-world scenario. Therefore, we need to verify
that high LRs we found with $S_F$ would hold even with larger number of possible authors.
For this purpose, we vary author set from $40$ to $1000$. In particular, we experiment
with a set of $[40, 100, 250, 500, 750, 1000]$ authors.
In each experiment, we assess the linkability of our
$40$ authors when mixing them with other authors.

Figure \ref{fig:training-authors} shows Top-1 and Top-4 LR of $S_F$ with $|AR|=5$.
Our preferred selection of features -- Top Letter Trigram and POS Bigrams --
achieves high linkability, $77.5\%$ in Top-1 and $90\%$ in Top-4, even in a set of $1000$ possible authors.

\subsection{Summary}
\label{sec:summary-analysis}
To summarize, our main results are as follows:
\begin{enumerate}
  \item We started with a well-known Writeprints feature set and achieved 
  modest LRs of up to $52.5\%$ in Top-1 and $82.5\%$ in Top-4 using the 
  CS model. (See Section \ref{itm:fs:1st})
  \item We then tried each Writeprint feature individually with the intuition 
  that the combination of multiple features would have more noise, thus decreasing
  linkability. Surprisingly, using only Top Letter Trigrams or POS Bigrams, we achieved
  significantly better LR than all Writeprints features.
  (See Section \ref{itm:fs:2nd})
  \item Next, we selected Top Letter Trigrams, which yields $91\%$ and $96\%$ LR in Top-1 and Top-4, as our rising main. 
  Then, we increased linkability to $96\%$ in Top-1, and $100\%$ in Top-4 by adding POS Bigrams.
  (See Section \ref{itm:fs:3rd})
  \item Even when assessing linkability within a large number of possible authors sets, the preferred combination 
  of features maintains high LR, e.g. $77.5\%$ in Top-1 and $90\%$ in Top-4 among $1000$ possible authors
  (See Section \ref{sec:multiauthors}). Thus, we end up setting $S_F$ as \{Top Letter Trigrams, POS Bigrams\},
  which will be used for evaluation of anonymization techniques.
\end{enumerate}

%% file: crowdsourcing.tex
\section{Fighting Authorship Linkability}
\label{sec:red-auth-link}
We now move on to the main goal of this paper: exploration of techniques
that mitigate authorship linkability. We consider two general approaches:  
\begin{enumerate}
\item Crowdsourcing: described in Section \ref{sec:crowdsourcing}.
\item Machine Translation: described in Section \ref{sec:translation}.
\end{enumerate}

\subsection{Crowdsourcing to the Rescue}
\label{sec:crowdsourcing}
We begin by considering what it might take, in principle, to anonymize reviews.
Ideally, an anonymized review would exhibit stylometric features that are not linkable,
with high accuracy, to any other review or a set thereof. At the same time, an anonymized
review must be as meaningful as the original review and must remain faithful or
 ``congruent'' to it. (We will come back to this issue later in the paper). 
We believe that such perfect anonymization is probably impossible.
This is because stylometry is not the only means of linking reviews. For example, 
if a {\sf TripAdvisor} contributor travels exclusively to Antarctica and her reviews
cover only specialized cruise-ship lines and related products (e.g., arctic-quality clothes),
then no anonymization technique can prevent linkability by topic without grossly distorting 
the original review. Similarly, temporal aspects of reviews might aid linkability\footnote{Here we mean 
time expressed (or referred to) within a review, not only time of posting of a review.}.
Therefore, we do not strive for perfect anonymization and instead confine the problem to 
the more manageable scope of reducing stylometric linkability. We believe that this 
degree of anonymization can be achieved by rewriting. 

\subsubsection{How to Rewrite Reviews?}
There are many ways of rewriting reviews 
in order to reduce stylometric linkability. One intuitive approach is to construct
a piece of software, e.g., a browser plug-in, that alerts the author about highly linkable
features in the prospective review. This could be done in real time, as the review
is being written, similarly to a spell-checker running in the background. Alternatively, 
the same check can be done once the review is fully written. The software might even proactively
recommend some changes, e.g., suggest synonyms, and partition long, or join short, sentences.
In general, this might be a viable and effective approach. However, we do not pursue it
in this paper, partly because of software complexity and partly due to the difficulty of
conducting sufficient experiments needed to evaluate it.

Our approach is based on a hypothesis
that the enormous power of global crowd-sourcing can be leveraged to efficiently  
rewrite large numbers of reviews, such that: 

{\bf (1)} Stylometric authorship linkability is appreciably reduced, and 

{\bf (2)} Resulting reviews remain sensible and faithful to the originals.

The rest of this section overviews crowdsourcing, describes 
our experimental setup and reports on the results.

\subsubsection{Crowdsourcing}
\label{sec:crowdsourcing-exp}
%
%
\noindent{\bf Definition:}
according to the Merriam-Webster dictionary, \texttt{Crowdsourcing} is defined as: 
{\em the practice of obtaining needed services, ideas, or content by soliciting contributions from 
a large group of people, and especially from an online community, rather than from 
traditional employees or suppliers.}

There are numerous crowdsourcing services ranging in size, scope and popularity.
Some are very topical, such as \url{kickstarter} (creative idea/project funding)
or \url{microworkers} (web site promotion), while others are fairly general, e.g.,
\url{taskrabbit} (off-line jobs) or \url{clickworker} (on-line tasks).
 
We selected the most popular and the largest general crowdsourcing service -- 
Amazon's Mechanical Turk (MTurk) \cite{amazonmturk}. This choice was made for several reasons:

\begin{itemize}
  \item We checked the types of on-going tasks in various general crowdsourcing services 
  and MTurk was the only one where we encountered numerous on-going text rewriting tasks. 

  \item We need solid API support in order to publish numerous rewriting tasks.
  We also need a stable and intuitive web interface, so that the crowdsourcing service can be easily
  used. Fortunately, MTurk has both a user-friendly web interface for isolated users and 
  API support to automate a larger number of tasks.

  \item Some recent research efforts have used MTurk for the purpose of similar studies
  \cite{passwords,mturk-example-2,mturk-example-3}.
\end{itemize}
In general, we need crowdsourcing for two phases: (1) rewrite original reviews,
and (2) conduct a readability and faithfulness evaluation between 
original and rewritten reviews. More than $400$ random 
MTurkers participated in both phases.

\subsubsection{Rewriting Phase}
\label{rewriting}
Out of 3 randomly created AR and IR review sets we used in Section \ref{sec:link-analysis},
we randomly selected one as the target for anonymization experiments.
We then uploaded all reviews in this AR set to the 
crowdsourcing service and asked MTurkers to rewrite them using their own 
words. We asked $5$ MTurkers to rewrite each review, in order
to obtain more comprehensive and randomized data for the subsequent linkability study. 
While rewriting, we explicitly instructed participants to keep the meaning similar and not to change 
proper names from the original review. Moreover, we checked whether the number of words in each
new review is close to that of the original before accepting a rewritten submission.
Divergent rewrites were
rejected\footnote{A sample rewriting task and its submission are shown in the Appendix of this paper's extended draft \cite{dropboxextended}.}.

We published reviews on a weekly basis in order to vary the speed
of gathering rewrites. Interestingly, most tasks were 
completed during the first 3 days of week, and the remaining 4 days 
were spent reviewing submissions. We finished the rewriting phase in 4 months. 
Given $40$ authors and AR size of $5$ ($200$ total original reviews), each review 
was rewritten by $5$ MTurkers, resulting in $1,000$ total submissions. Of these, 
we accepted $882$. The rest were too short or too long, not meaningful, not faithful
enough, or too similar, to the original. Moreover, out of $200$ originals, 
$139$ were rewritten $5$ times. All original and rewritten reviews can be 
found at our publically shared folder \cite{dropboxdataset}.

We paid US$\$0.12$, on average, for each rewriting task. 
Ideally, a crowdsourcing-based review rewriting system would be free, with peer
reviewers writing their own reviews and helping to re-writing others.
However, since there was no such luxury at our disposal, 
we decided to settle on a low-cost approach\footnote{We consider the average of
US$\$0.12$ to be very low per review cost.}.
Initially, we offered to pay US$\$0.10$ per rewritten review. 
However, because review size ranges between $2$ and $892$ words, 
we came up with a sliding-price formula: $\$0.10$ for every $250$ 
words or a fraction thereof, e.g., a $490$-word review pays $\$0.20$ 
while a $180$-word one pays $\$0.10$. In addition, Amazon MTurk 
charges a $10\%$ fee for each task.

One of our secondary goals was assessment of efficacy and usability of the 
crowdsourcing service itself. We published one set of $40$ reviews via the 
user interface on the MTurk website, and the second set of $160$ reviews -- using 
MTurk API. We found both means to be practical, error-free and easy to use. 
Overall, anyone capable of using a web browser can easily publish their reviews on
MTurk for rewriting.

After completing the rewriting phase, we continued with a readability study
to assess sensibility of rewritten reviews and their correspondence to the originals.

\subsubsection{Readability Study}
\label{readability}
Readability study proceeded as follows: First, we pick, at random, $100$ reviews
from $200$ reviews in the AR set. Then, for each review, 
we randomly select one rewritten version. Next, for every [original,rewritten] 
review-pair, we publish a readability task on MTurk. In those tasks, we ask two distinct
MTurkers to score rewritten reviews by comparing its similarity and sensibility to the original one.
We define the scores as Poor(1), Fair(2), Average(3), Good(4), Excellent(5), 
where Poor means that the two reviews are completely different, and Excellent means they
are essentially the same meaning-wise. We also ask MTurkers to write a comprehensive result
which explains the differences (if any) between original and rewritten
counterparts\footnote{A sample readability study task and its submission are presented in the
Appendix of this paper's extended draft \cite{dropboxextended}.}.

\begin{figure}[H]
  \centering
  \includegraphics[scale=0.50]{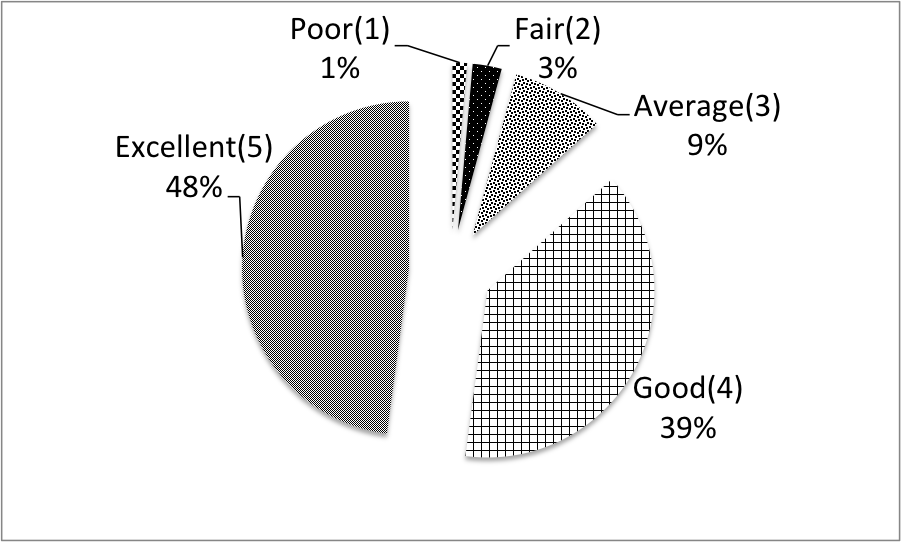}
  \caption{Readability Results of Rewritten Reviews}
  \label{fig:readability_rewritten}
\end{figure}

This study took one week and yielded $142$ valid submissions.
Results are reflected in Figure \ref{fig:readability_rewritten}.
The average readability score turns out to be $4.29/5$, while $87\%$ of reviews are given scores of
Good or Excellent. This shows that rewritten reviews generally retain the meaning of the originals.
Next, we proceed to re-assess stylometric linkability of rewritten reviews.

\begin{figure*}[t]
  \centering
  \subfigure[Top-1 while varying the size of author set]{\label{fig:original_vs_rewritten_top1_authors-png}
  \includegraphics[scale=0.42]{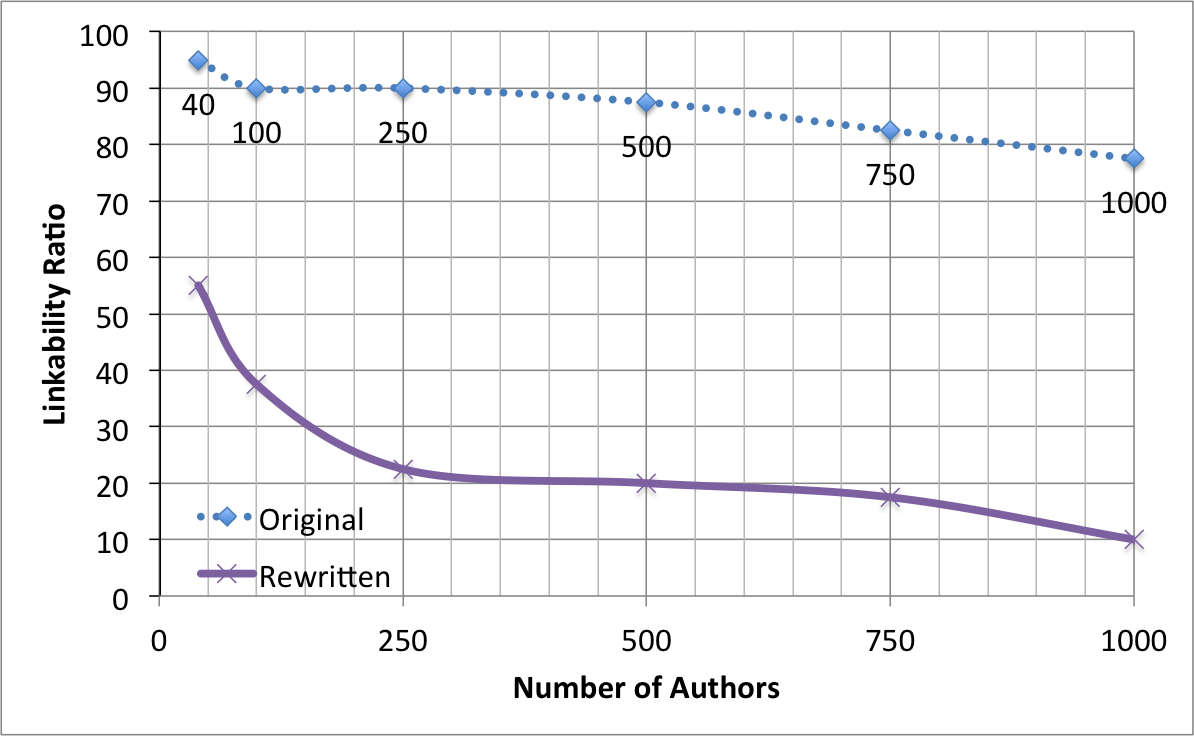}}
  \subfigure[Top-4 while varying the size of author set]{\label{fig:original_vs_rewritten_top4_authors-png}
  \includegraphics[scale=0.42]{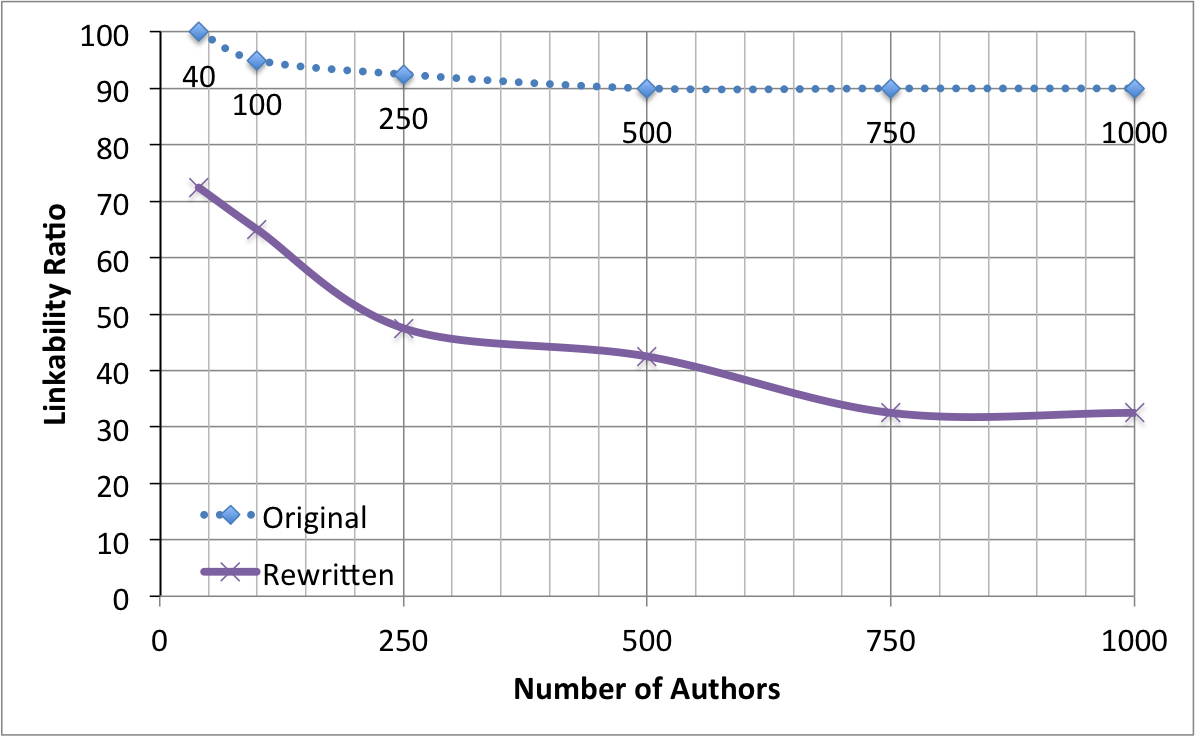}}
  \subfigure[Top-1 in a set of 1000 authors]{\label{fig:original_vs_rewritten_top1-png}
  \includegraphics[scale=0.46]{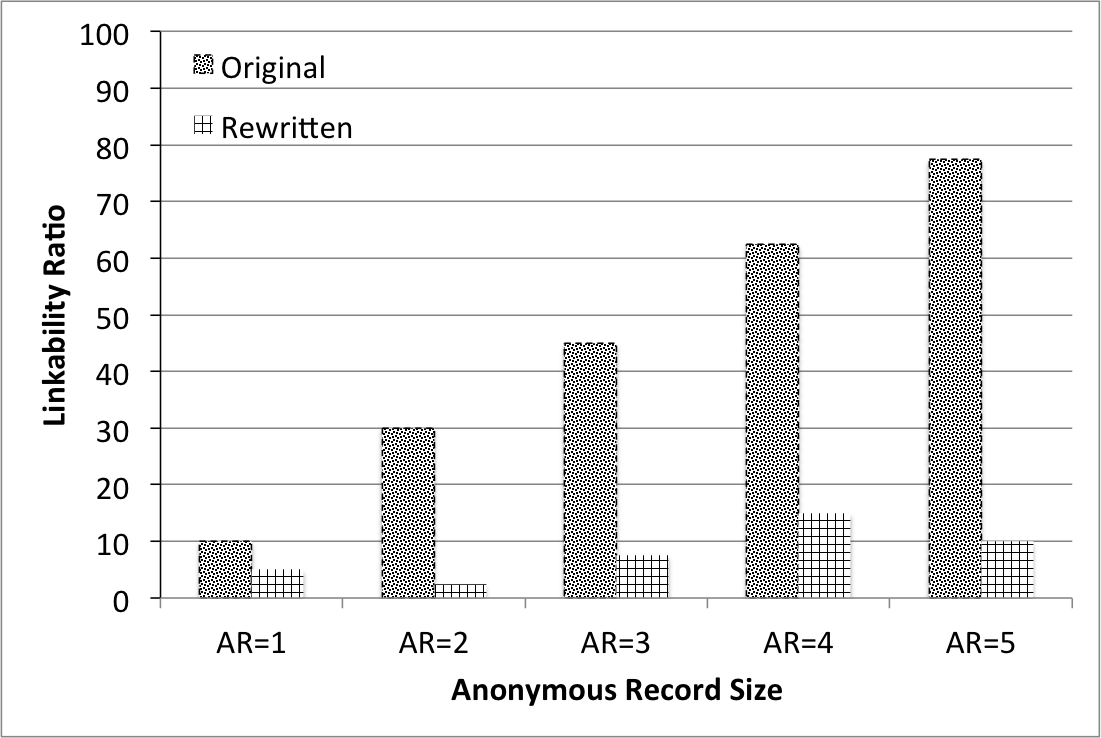}}
  \subfigure[Top-4 in a set of 1000 authors]{\label{fig:original_vs_rewritten_top4-png}
  \includegraphics[scale=0.46]{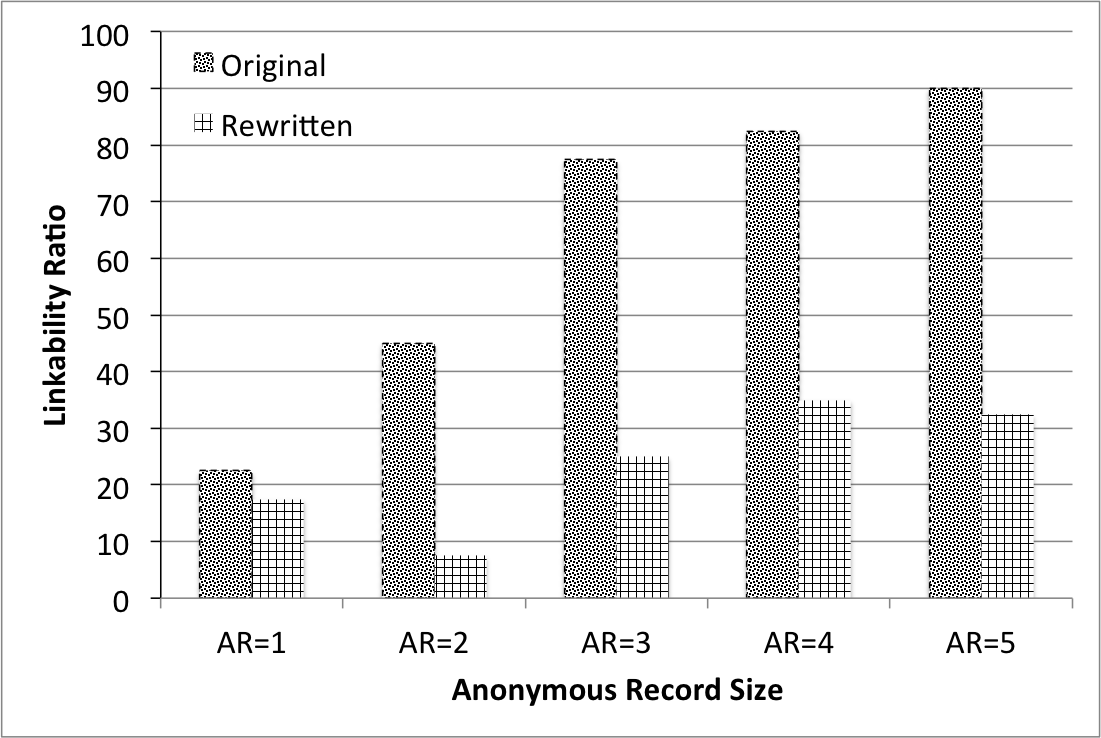}}
  \caption{LRs of Original and Rewritten Reviews}
  \label{fig:original-vs-rewritten}
\end{figure*}

\subsubsection{Linkability of Rewritten Reviews}
\label{linkability_rewritten}
Recall that the study in Section \ref{sec:link-analysis} involved $3$ review
sets each with $100$ reviews per author. For the present study, we only consider the first
set since we published anonymous reviews from first set to MTurk. In this first set, we replace AR with the 
corresponding set of MTurk-rewritten reviews where we pick a random rewritten version of each review,
while each author's IR remains the same. 

Figures \ref{fig:original_vs_rewritten_top1_authors-png} and \ref{fig:original_vs_rewritten_top4_authors-png}
compare LRs between original-rewritten reviews with varying number of authors.
Interestingly, we notice a substantial decrease in LRs for all author sizes.
For $|AR|=5$ in a set of $1000$ authors, Top-1 and Top-4 LR drop from $77.5\%$ to $10\%$ and from $90\%$ to $32.5\%$ respectively.
Even only in $40$ authors set, Top-1 LR decreases to $55\%$, which is significantly lower than $95\%$ achieved with original reviews.

We also present a detailed comparison of original and rewritten reviews' LRs with different AR sizes
in Figures \ref{fig:original_vs_rewritten_top1-png} and \ref{fig:original_vs_rewritten_top4-png}.
Notably, both Top-1 and Top-4 LR decrease dramatically for all AR sizes. $35\%$
is the highest LR obtained with rewritten reviews, which is substantially less than those achieved with original counterparts.

After experiencing this significant decrease in linkability, we analyze
rewritten reviews to see what might have helped increase anonymity.
We notice that most MTurkers do not change the skeleton of original review.
Instead, they change the structure of individual sentences
by modifying the order of subject, noun and verb, converting an active sentence into
a passive one, or vice versa. We also observe that MTurkers swap words with synonyms.
We believe that these findings can be combined into an automated tool, which can help authors rewrite
their own reviews. This is one of the items for future work, discussed in more detail in Section \ref{sec:discussion}.

\subsubsection{Crowdsourcing Summary}
\label{sec:crowdsourcing-summary}
We now summarize key findings from the crowdsourcing experiment.
\begin{enumerate}
  \item MTurk based crowdsourcing yielded rewritten reviews that were: 
    \begin{compactitem}
      \item [\bf Low-cost] -- we paid only $\$0.12$ including 10\% service fee for rewriting each $250$-word review.
      \item [\bf Fast] -- we received submissions within 3-4 days, on average.
      \item [\bf Easy-to-use] -- based on experiences with both user-interface and API of MTurk,
      an average person who is comfortable using a browser, Facebook or Yelp can easily publish 
      reviews to MTurk.
    \end{compactitem}

  \item As the readability study shows, crowdsourcing produces meaningful results: rewrites 
  remain faithful to originals. (See Section \ref{readability}).

  \item Most importantly, rewrites substantially reduce linkability. For an $|AR|=5$
  where we previously witnessed the highest LR, Top-1 LR shrunk from $95\%$ to $55\%$ in a set of $40$ authors and 
  from $77.5\%$ to $10\%$ in a set of $1000$ authors.  (See Section \ref{linkability_rewritten}). 
\end{enumerate}

%% file: translation.tex
\subsection{Translation Experiments}
\label{sec:translation}
We now discuss an alternative approach that uses on-line translation
to mitigate linkability discussed in Section \ref{sec:link-analysis}.
The goal is to assess the efficacy of translation for stylometric obfuscation
and check whether, in combination with crowdsourcing, it can be blended into a single 
socio-technological linkability mitigation technique.

It is both natural and intuitive to consider machine (automated, on-line) translation 
for obfuscating stylometric features of reviews. 
One well-known technique is to incrementally translate the text into a sequence 
of languages and then translate back to the original language. 
For example, translating a review from (and to) English
using three levels of translation (two intermediate languages) could be done as follows:
English $\rightarrow$ German $\rightarrow$ Japanese $\rightarrow$ English. The main intuition 
is to use the on-line translator as an external re-writer, so that stylometric characteristics 
would change as the translator introduces its own characteristics. 

Using a translator to anonymize writing style has been attempted in prior work
\cite{brennan2012adversarial,translate-once-twice}. However, 
prior studies did not go beyond three levels of translation and 
did not show significant decreases in linkability. Also, it was shown that   
that translation often yields non-sensical results, quite divergent from the original text \cite{brennan2012adversarial}.
Due to recent advances in this area, we revisit and reexamine the use of translation. 
Specifically, we explore effects of the number of intermediate languages on linkability and 
assess readability of translated outputs. In the process, we discover 
that translators are actually effective in mitigating linkability,
while readability is (though not great) is reasonable and can be easily fixed by crowdsourcing.  

\subsubsection{Translation Framework}
We begin by building a translation framework to perform a large number of 
translations using any number of languages. Currently, Google \cite{googletranslate} and 
Bing \cite{bingtranslate} offer the most popular machine translation services.
Both use statistical machine translation techniques to dynamically translate 
text between thousands of language pairs. Therefore, given the same text, they usually return a 
different translated version. Even though there are no significant differences between them,
we decided to use Google Translator. It supports more languages: 64 at the time of this writing
\cite{googletranslatorapi}, while Bing supports 41 \cite{bingtranslatorapi}).

Google provides a translation API as a free service to researchers with a daily character quota,
which can be increased upon request. The API provides the following two functions:
\begin{compactitem}
  \item $translate(text, sourceLanguage, targetLanguage)$: Translates given text from source language to target language.
  \item $languages()$: Returns the set of source and target languages supported in the \textit{translate} function.
\end{compactitem}
Using these functions, we implement the algorithm, shown in Algorithm \ref{translation-algo}.
We first select $N$ languages at random. Then, we
consecutively translate text into each of the languages, one after the other. 
At the end, we translate the result to its original language, English, in our case. 
We consider the final translated review as the anonymized version of the original. 

We also could have used a fixed list of destination languages. However, it is easy to see that translated reviews 
might then retain some stylometric features of the original (This is somewhat analogous to deterministic encryption.).
Thus, we randomize the list of languages hoping that it would make it improbable to retain stylometric patterns.
For example, since Google translator supports $64$ languages, we have more than 
$\prod\limits_{n=0}^{N-1} (64-n) \approx 2^{53}$ distinct lists of languages for $N=9$.  
\begin{algorithm}
\caption{Round-Translation of $Review$ with $N$ random languages}
\begin{algorithmic}
\STATE Obtain all supported languages via $languages()$
\STATE $RandomLanguages \leftarrow $ select $N$ languages randomly
\STATE $Source \leftarrow $ ``English''
\FOR{Language $language$ in $RandomLanguages$}
  \STATE $Review \leftarrow$ translate($Review$, $Source$, $language$)
  \STATE $Source \leftarrow language$
\ENDFOR
\STATE $Translated \leftarrow$ translate($Review$, $Source$, ``English'')
\RETURN{$Translated$}
\end{algorithmic}
\label{translation-algo}
\end{algorithm}

After implementing the translation framework, we proceed to
assessing linkability of the results.

\begin{figure*}[t]
  \centering
  \subfigure[Comparison of Original-Translated Reviews in Top-1]
    {\label{fig:original_vs_translation_top1-png}
    \includegraphics[scale=0.44]{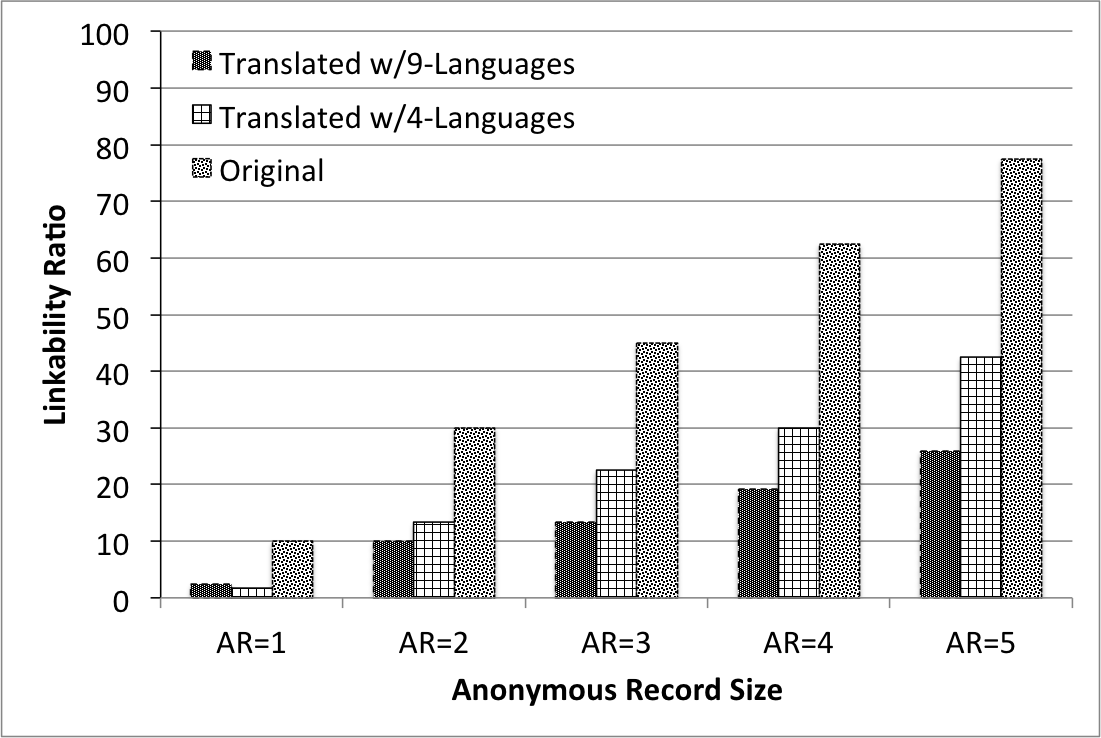}}
  \subfigure[Comparison of Original-Translated Reviews in Top-4]
    {\label{fig:original_vs_translation_top4-png}
    \includegraphics[scale=0.44]{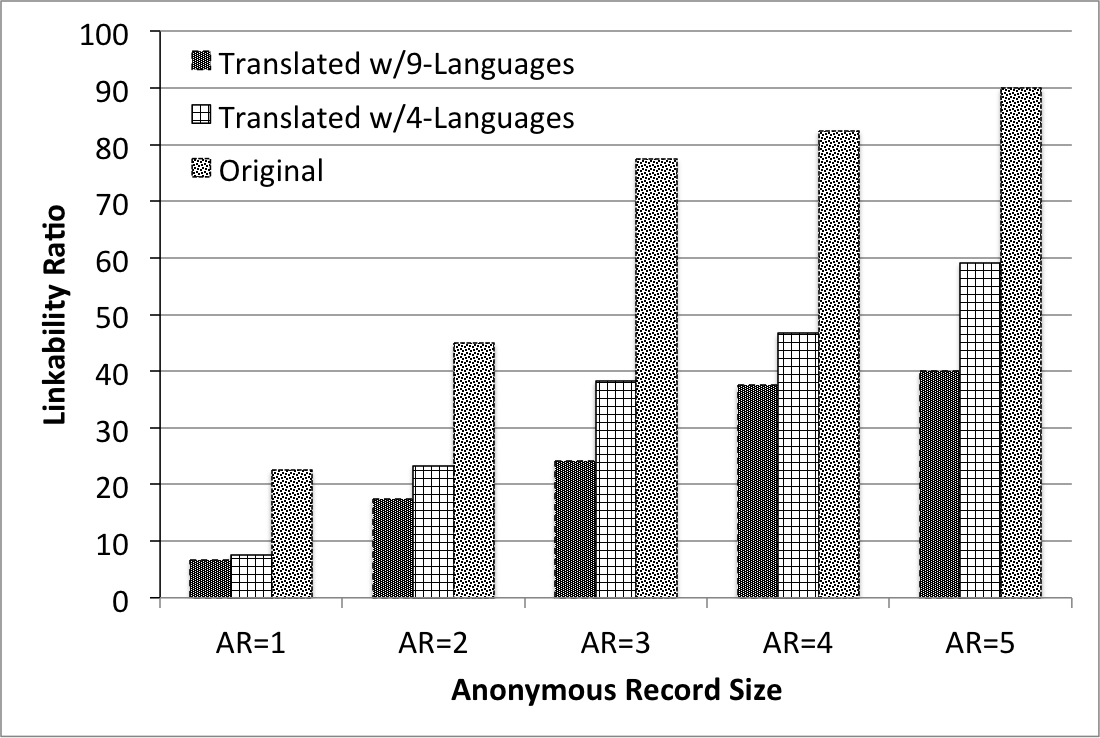}}
  \subfigure[LRs with $|AR|=5$ while varying number of languages]
    {\label{fig:translation_comparison-png}
    \includegraphics[scale=0.44]{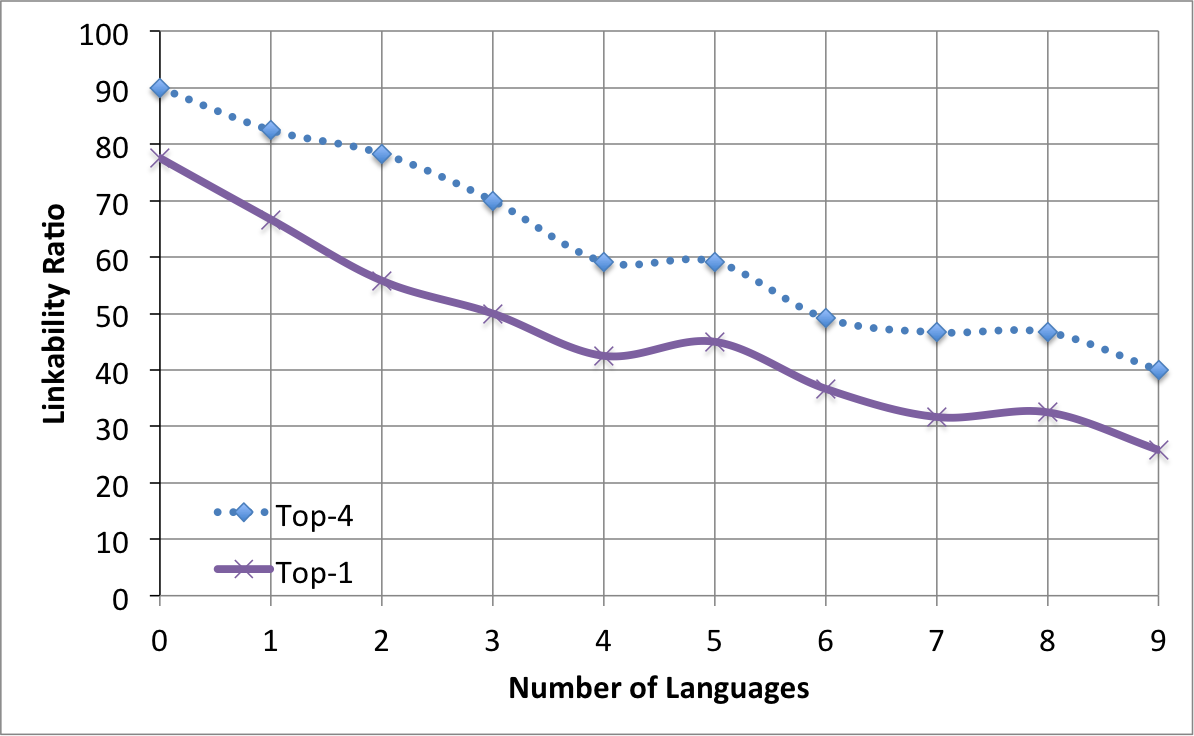}}
  \caption{LRs of Translated Reviews in a set of $1000$ authors}
  \label{fig:translated-linkability-png}
\end{figure*}

\subsubsection{Linkability of Translated Reviews}
Using Algorithm \ref{translation-algo}, we anonymized the AR review
set\footnote{Translated example reviews are shown in the Appendix of this paper's extended draft \cite{dropboxextended}.}.
We varied $N$ from $1$ to $9$ and re-ran linkability analysis with translated reviews as the AR.
In doing so, we used $S_F$ identified in Section \ref{sec:link-analysis}. To assert 
generality of linkability of translated texts, we performed the above procedure
$3$ times, each time with a different list of random languages and then ran linkability
analysis $3$ times as well. Average linkability results of all $3$ runs are plotted
in Figures \ref{fig:original_vs_translation_top1-png}, \ref{fig:original_vs_translation_top4-png}
and \ref{fig:translation_comparison-png}.

For the number of intermediate languages, our intuition is that 
increasing the number of levels of translation (i.e., intermediate languages)
causes greater changes in stylometric characteristics of original text. 
Interestingly, Figure \ref{fig:translation_comparison-png} supports this intuition: larger $N$ values 
yield larger decreases of linkability. While the decrease is not significant in Top-4 for $N$: $[1, 2]$, it becomes more 
noticeable after $3$ languages. For $|AR|=5$, we have Top-1 \& Top-4 linkabilities of $42.5\%$ \& $59\%$ with $4$ languages,
$31\%$ \& $47\%$ with $7$ languages and $25\%$ \& $40\%$ with $9$ languages, respectively. These
are considerably lower than $77.5\%$ \& $90\%$ achieved with original ARs.
Because Top-1 linkability decreases to $25\%$ after $9$ languages, we stop increasing $N$ and settle on $9$.

Figures \ref{fig:original_vs_translation_top1-png} and \ref{fig:original_vs_translation_top4-png}
show reduction in Top-1 and Top-4 linkability for varying AR sizes.
In all of them, original reviews have higher LRs than ones translated with $4$ languages; which in turn
have higher LRs than those translated with $9$ languages. This clearly demonstrates that
when more translations are done, the more translator manipulates the stylometric characteristics of a review.

\subsubsection{Readability of Translated Reviews}
\label{trans:readability} 
So far, we analyzed the impact of using on-line translation on decreasing stylometric linkability. 
However, we need to make sure that the final result is readable.
To this end, we conducted a readability study. We randomly selected a sample of translated reviews for $N=9$.
We have 3 sets of translated reviews, each corresponding to a random selection of 9 languages. 
From each set, we randomly selected 20 translated reviews, which totals up to 60 translated reviews.
Then, for each $[original,translated]$ review-pair, we published readability tasks on MTurk
(as in Section \ref{readability}) and had it assessed by 2 distinct MTurkers, resulting in 120 total submissions.

\begin{figure}[H]
  \centering
  \includegraphics[scale=0.50]{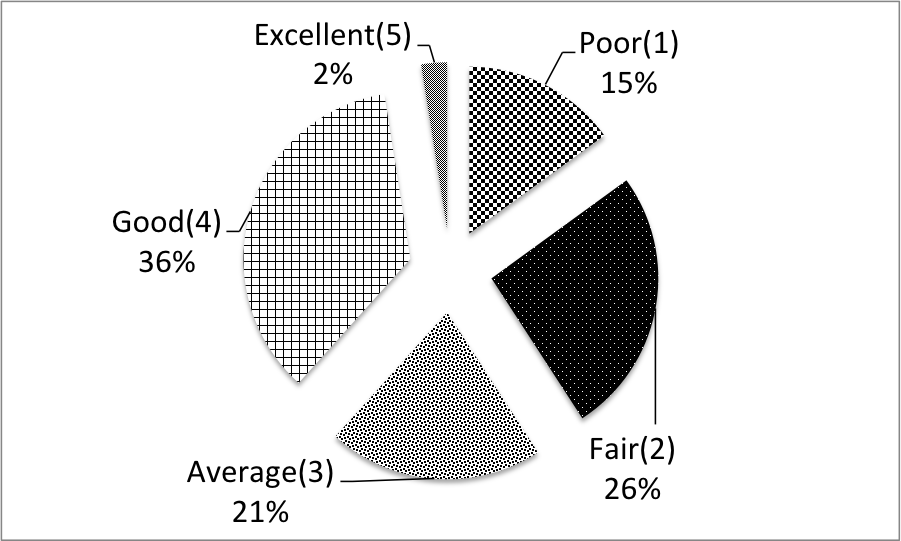}
  \caption{Readability Results of Translated Reviews}
  \label{fig:readability_translated}
\end{figure}

Results are shown in Figure \ref{fig:readability_translated}.
As expected, results are not as good as those in Section \ref{readability}.
However, a number of reviews preserve the original meaning to some extent. 
The average score is $2.85$ out of $5$ and most scores were at least ``Fair". 



\subsubsection{Fixing the Translated Reviews}
Even though machine translation is continuously getting better at producing readable output, 
the state-of-the-art is far from ideal. After manually reviewing some $[original,translated]$ pairs, 
we realized that most translated reviews retained the main idea of the original. However, because of:
(1) frequently weird translation of proper nouns, (2) mis-organization of sentences, and (3) failure of 
translating terms not in the dictionary, translated review are not easy to read. We decided to provide 
translated reviews along with their original versions to MTurkers and asked them to fix 
unreadable parts\footnote{A sample translation-fix task and its submission
are presented in the Appendix of this paper's extended draft \cite{dropboxextended}.}.
As a task, this is easier and less time-consuming than rewriting the entire review.

\begin{figure*}[t]
  \centering
  \subfigure[LRs in a set of $1000$ authors while varying the AR size]{\label{fig:31authors_comparison_top1-png}
  \includegraphics[scale=0.42]{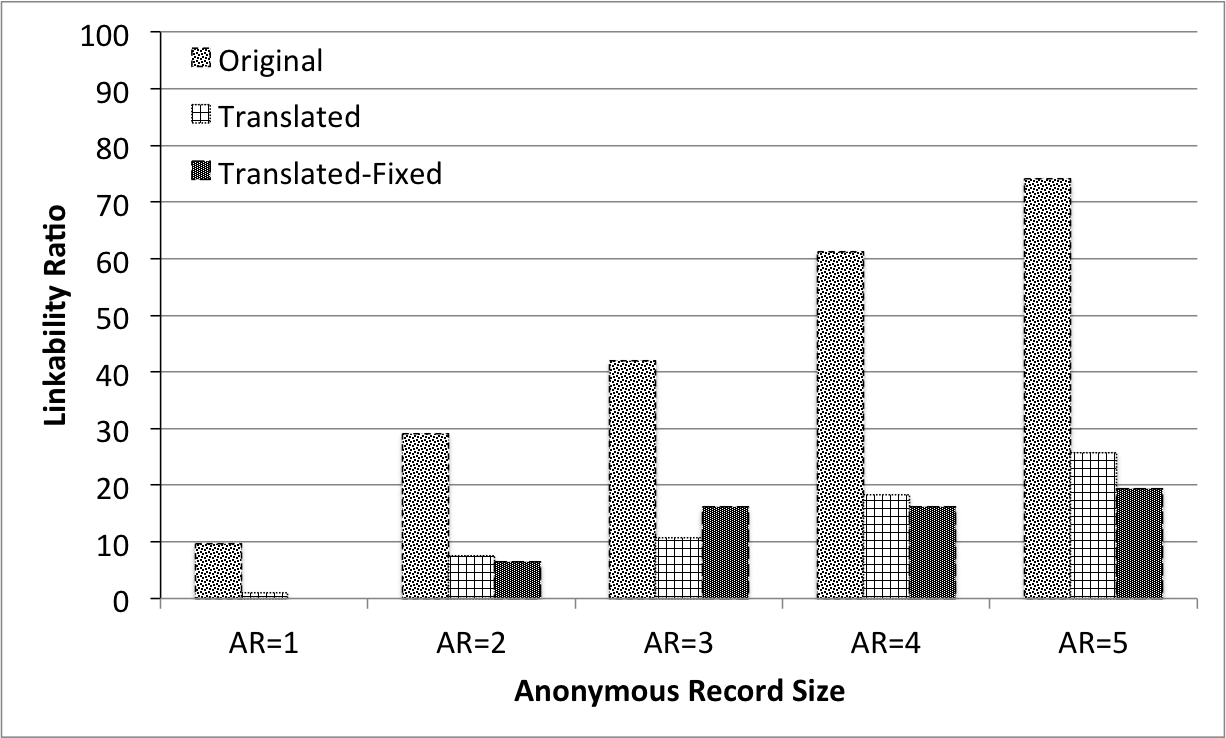}}
  \subfigure[LRs with $|AR|=5$ while varying the number of authors]{\label{fig:31authors_comparison_top1_authors-png}
  \includegraphics[scale=0.42]{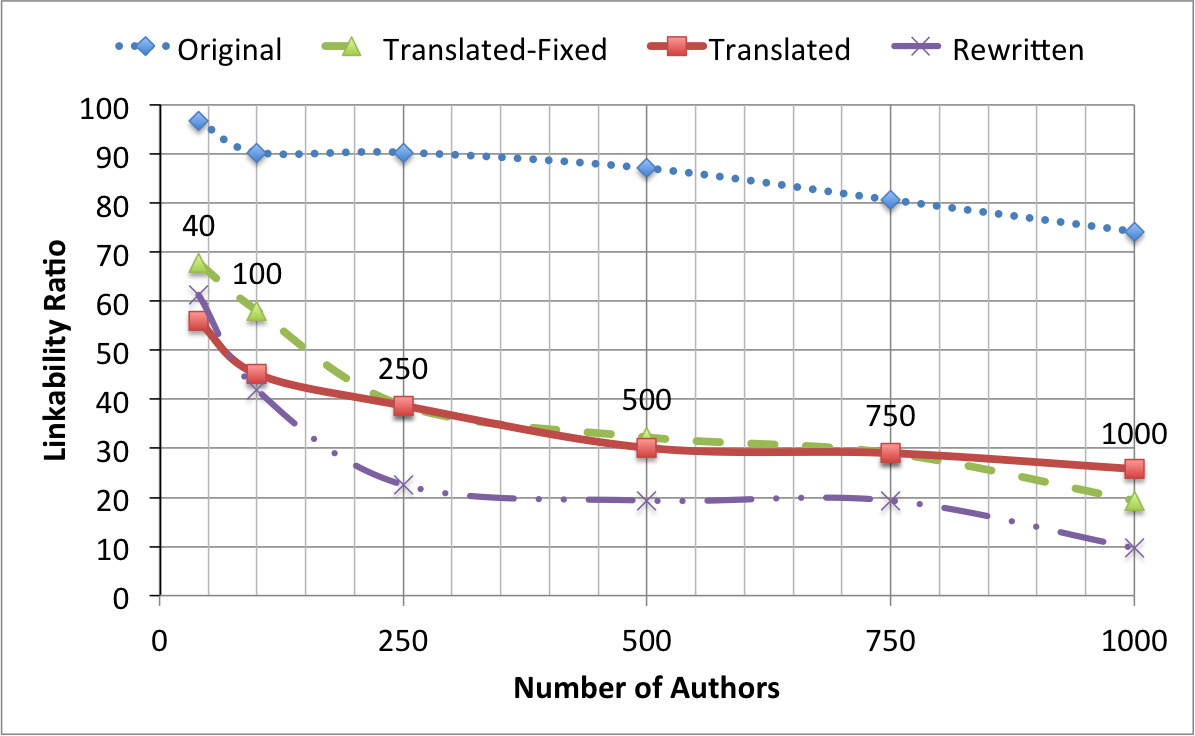}}
  \caption{Top-1 LRs for Original, Translated, Translation-Fixed and Rewritten Reviews}
  \label{fig:31authors_comparison}
\end{figure*}

Out of $3$ translated review sets, 
we selected one at random and published all $200$ ($|AR|=5$ for 40 authors) 
translated reviews from our AR set to MTurk. We received $189$ submissions; only $31$ authors had their
full AR's translated reviews completely fixed. We then performed the same linkability assessment with
these $31$ authors while we update their AR's by translated-fixed reviews.

Comparison of linkability ratios between original, translated, and fixed version of the same translated
reviews is plotted in Figure \ref{fig:31authors_comparison_top1-png}. It demonstrates that,
fixing translations does not significantly influence linkability. In AR-5, Top-1 linkability
of fixed translation is $19\%$ while non-fixed translations $25\%$.
Meanwhile, both are significantly lower than $74\%$ LR of original counterparts.

\begin{figure}[H]
  \centering
  \includegraphics[scale=0.45]{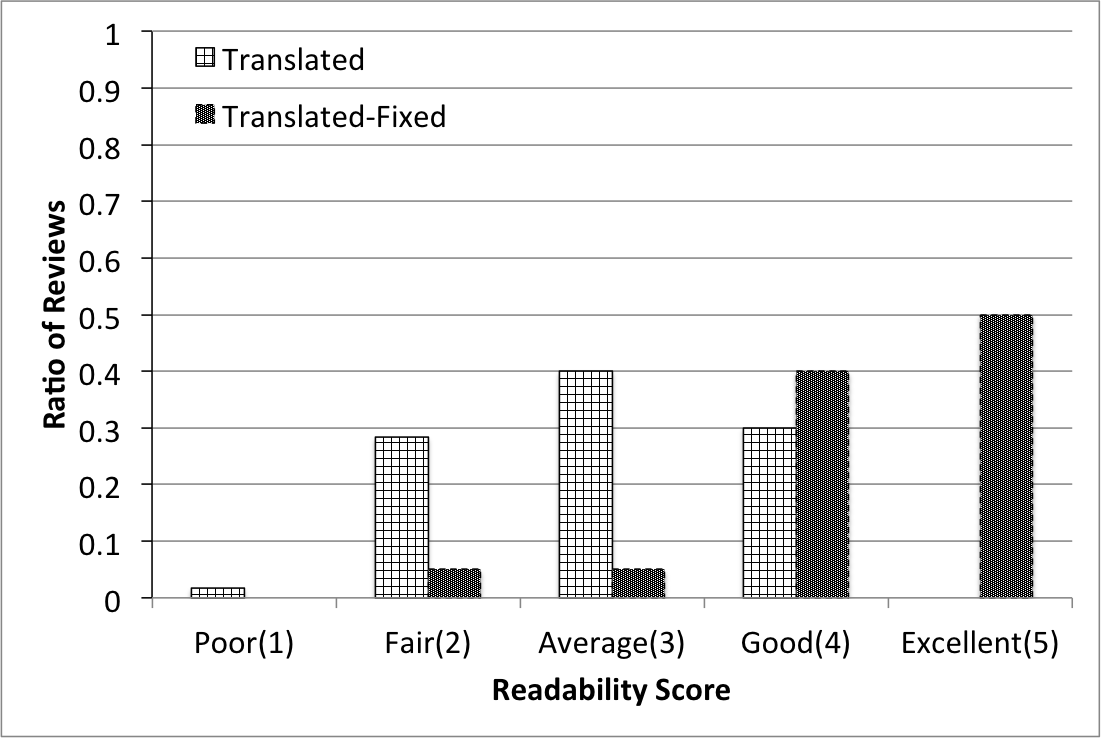}
  \caption{Readability study comparison between Translated and Translated-Fixed Reviews}
  \label{fig:readability_translated_fixed}
\end{figure}

Finally, we perform a readability study on fixed translations. Out of $189$ submissions,
we select $20$ randomly and publish to MTurk as a readability task. Average readability
score increased from $2.85$ to $4.12$ after fixing the machine translation.
Detailed comparison of readability studies between translated and translated-fixed
reviews is given in Figure \ref{fig:readability_translated_fixed}. We notice high percentage of
translated reviews has Average score, while fixed counterparts mostly score as Good or Excellent.
Results are really promising since they show that the meaning of a machine translated review can be fixed
while keeping it unlinkable. 

\subsubsection{Comparison of Anonymization Techniques}
We present the comparison of linkability results achieved using crowdsourcing,
machine translation and combination of both in Figure \ref{fig:31authors_comparison_top1_authors-png}.
Regardless of the size of author set, we achieve substantial decrease
in linkability. Our techniques show that
people are good at rewriting and correcting reviews while introducing their own style,
keeping the meaning similar, and, most importantly, reducing linkability.
While purely rewritten reviews have the lowest linkability, both
translated and translated-fixed reviews perform comparable to each other.
As far as readability, crowdsourcing (mean score of $4.29/5$)
performed much better than translation (mean score of $2.85/5$).
However, results show that low readability scores can be fixed
(resulting in a mean score of $4.12/5$) using crowdsourcing while keeping linkability low.
We summarize results as follows:

\begin{compactitem}
  \item Crowdsourcing:
  Achieves better anonymity and readability. However, it takes longer than translation
  since it is not an automated solution. Moreover, though not expensive, it is
  clearly not free.
  \item Machine Translation:
  Completely automated and cost-free approach which takes less time than crowdsourcing. 
  However, poor readability is the main disadvantage.
\end{compactitem}

%% file: discussion.tex
\section{Discussion}
\label{sec:discussion}
In spite of its usefulness in decreasing linkability and enhancing readability,
there are some open questions associated with the use of crowdsourcing.

\begin{enumerate}
\item \textbf{How applicable is crowdsourcing to other OSNs?}
In some other OSNs penalties for deanonymization would be higher
than Yelp. However we chose Yelp dataset for the reasons given in Section \ref{sec:dataset}.
The same technique can be presumably applied to other settings, e.g.,
anonymous activist blogs, tweets in Twitter and TripAdvisor reviews.

\item \textbf{How might authors get their reviews rewritten?}
This could be addressed by integrating a plug-in into a browser. When an author
visits an OSN and writes a review, this plug-in can ease posting of a task to a crowdsourcing platform and return
the result back to the author via one-time or temporary email address. On the system side, plug-in would create
a rewriting task and relay it to the crowdsourcing system.
A possible building block can be the recent work in \cite{mturk-automate}
that proposes a crowdsourcing task automation system. It automates task scheduling, pricing
and quality control, and allows tasks to be incorporated into the
system as a function call.

\item \textbf{How feasible is crowdsourcing in terms of latency and cost?}
We believe that a delay of couple of days would not pose an inconvenience
since review posting does not need to occur in real time.
Many popular OSNs does not publish reviews instantly, e.g., TripAdvisor screens each review
to make sure it meets certain guidelines. This moderation can take as long as several weeks \cite{tripadvisor}.

As far as costs, we paid US\$$0.12$, on average for each rewriting task. We consider
this amount is extremely low which can be easily subsidized by the advertizing revenue, with ads in the plug-in.


\item \textbf{Is there a privacy risk in posting reviews to strangers?}
It is difficult to assess whether there is a privacy risk since an adversary does not learn both posted
and rewritten reviews, unless she is registered as a worker, completes the task, and
her submission gets published. However, this clearly does not scale for the adversary when the number
of posted reviews is large and requires manual follow-up with the posts. Also,
MTurk Participation Agreement\footnote{https://www.mturk.com/mturk/conditionsofuse}
involves conditions that protect privacy of both worker and requester.

\item \textbf{Is there a chance of having a rewriter's writing style recognized?}
We believe that this is not the case.
First, there are many workers to choose from and we can force the system not to select the
same worker more than a specific number of times.
Second, we expect that a worker would rewrite many reviews from different sources.
This will widen the range of topics that rewritten reviews would cover and would make
rewritten reviews more difficult to recognize.
Finally, the identities of workers are expected to remain private since the only
party who can see worker details for a given task is the person who posted it.

\end{enumerate}

%% file: conclusion.tex
\section{Conclusions and Future Work}
\label{sec:conclusion}
This paper investigated authorship linkability in community reviewing and 
explored some means of mitigating it.
First, we showed, via a linkability study using a proper subset of the Writeprints feature set,
that authorship linkability is higher than previously reported. Then, using the 
power of global crowdsourcing on the Amazon MTurk platform, we published reviews and
asked random strangers to rewrite them for a nominal fee. After that, we conducted a 
readability study showing that rewritten reviews are meaningful and remain similar 
to the originals. Then, we re-assessed linkability of rewritten 
reviews and discovered that it decreases substantially.  Next, we considered
using translation to rewrite reviews and showed that linkability decreases
while number of intermediary languages increases. After that, we evaluated 
readability of translated reviews,
and realized that on-line translation does not yield results as readable as those from 
rewritings. Next, we take advantage of crowdsourcing to fix poorly readable translations
and still achieve low linkability.

This line of work is far from being complete and many issues remain
for future consideration:
\begin{compactitem}
%
\item We need to explore detailed and sophisticated evaluation techniques in order to 
understand stylometric differences between original, rewritten and translated reviews. 
If this succeeds, more practical recommendations can be given to review authors.
\item As discussed in Section \ref{sec:discussion}, we want to parlay the results
of our study into a piece of software or a plug-in intended for authors. 

\item We need to conduct the same kind of study in the context of 
review sites other than Yelp, e.g., Amazon, TripAdvisor or Ebay.
Also, cross-site studies should be undertaken, e.g., using a combination of 
Amazon and Yelp reviews. 
%
%
%
\end{compactitem}

%% file: appendix.tex
\section*{Appendix A: Crowdsourcing Examples}
\label{sec:crowdsourcing-example}
We present two example submissions from our rewriting and readability tasks in 
MTurk. Note that the full collection of original and rewritten reviews can be accessed 
in \cite{dropboxdataset}.

\subsection*{A.1: Rewriting Example}
\label{sec:crowdsourcing-rewriting}
Sample rewritten review for the task given in 
Figure ~\ref{fig:mt-rewrite-exp-png}:
\noindent{\textit{``When arriving the line was all the way around the block, 
so we were more than willing to sit with strangers. This wasn't what bothered 
me the most. What bothered me the most was that we were seated way in the 
back of the establishment. When the cart pushers bothered to help us they 
had no more food left except for chicken feet. A half an hour went by before 
we got the attention of staff to let them know that we needed to be fed. 
The food was delicious, or it was my hunger that stimulated my appetite. 
I chose eggplant, which was better at China Garden. However, Sino-American 
women with their mothers came in here. The lady delivering food had a 
consistent smug look on her face. Authentic and super.''}}

\subsection*{A.2 Readability Example}
\label{app:crowdsourcing-readability}

Sample readability study submission for a task given in
Figure \ref{fig:example-readability-box-png}:

\noindent{\textit{``Good, they are similar on the main points. When discussing the 
Sino-American women and their mothers, the alternative left out `pretty' which 
seemed like a point that appealed to original reviewer. Also, by using the word 
`smug' to describe the lady delivering food, the alternative review 
depicts her more negatively than the original.''}}

\section*{Appendix B: Translation Examples}
\label{app:translation-examples}
\begin{figure*}[t]
  \centering
  \subfigure[Sample rewriting task in MTurk]{\label{fig:mt-rewrite-exp-png}
  \fbox{\includegraphics[scale=0.25]{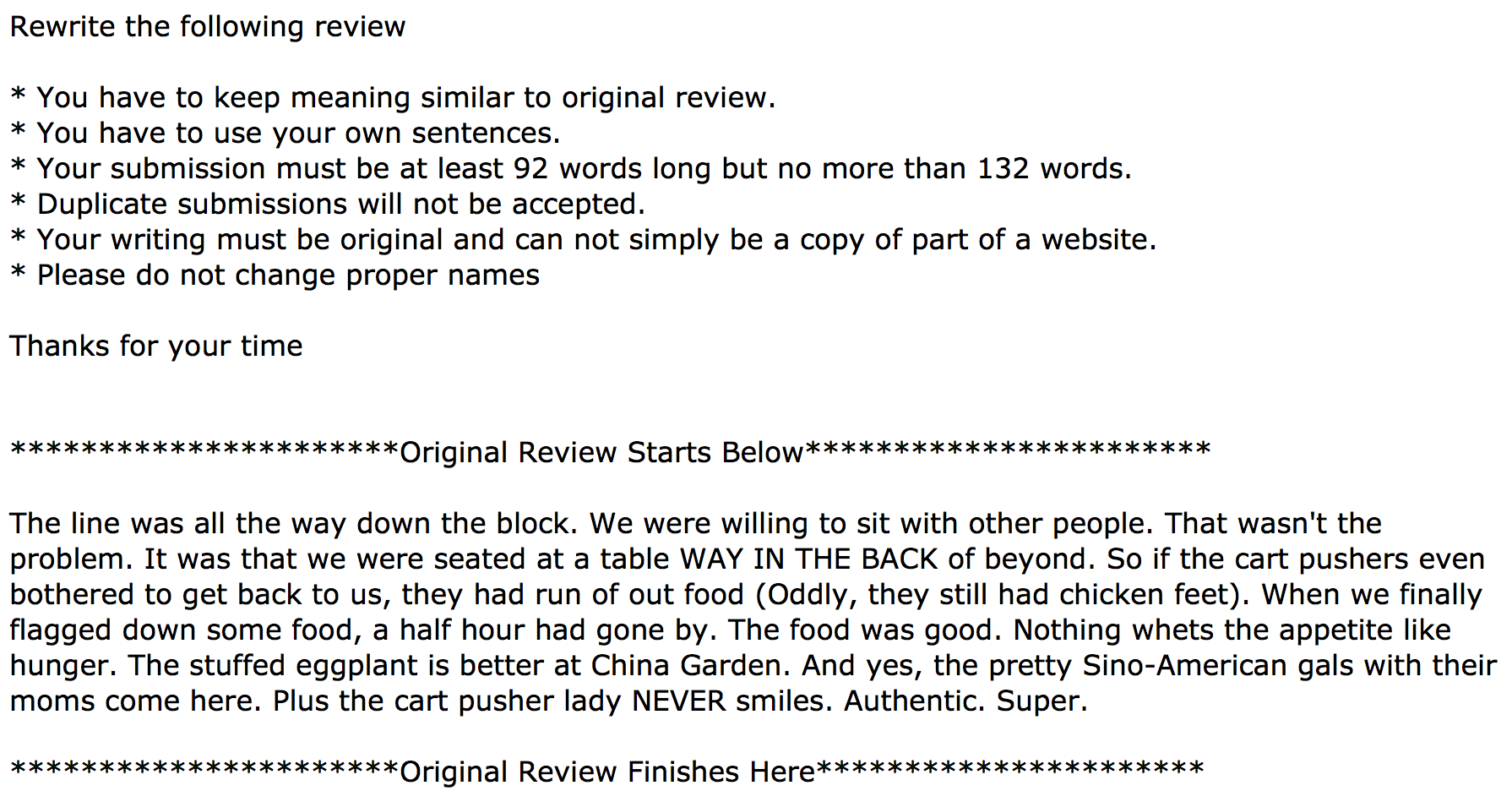}}}
  \subfigure[Sample readability task in MTurk]{\label{fig:example-readability-box-png}
  \fbox{\includegraphics[scale=0.25]{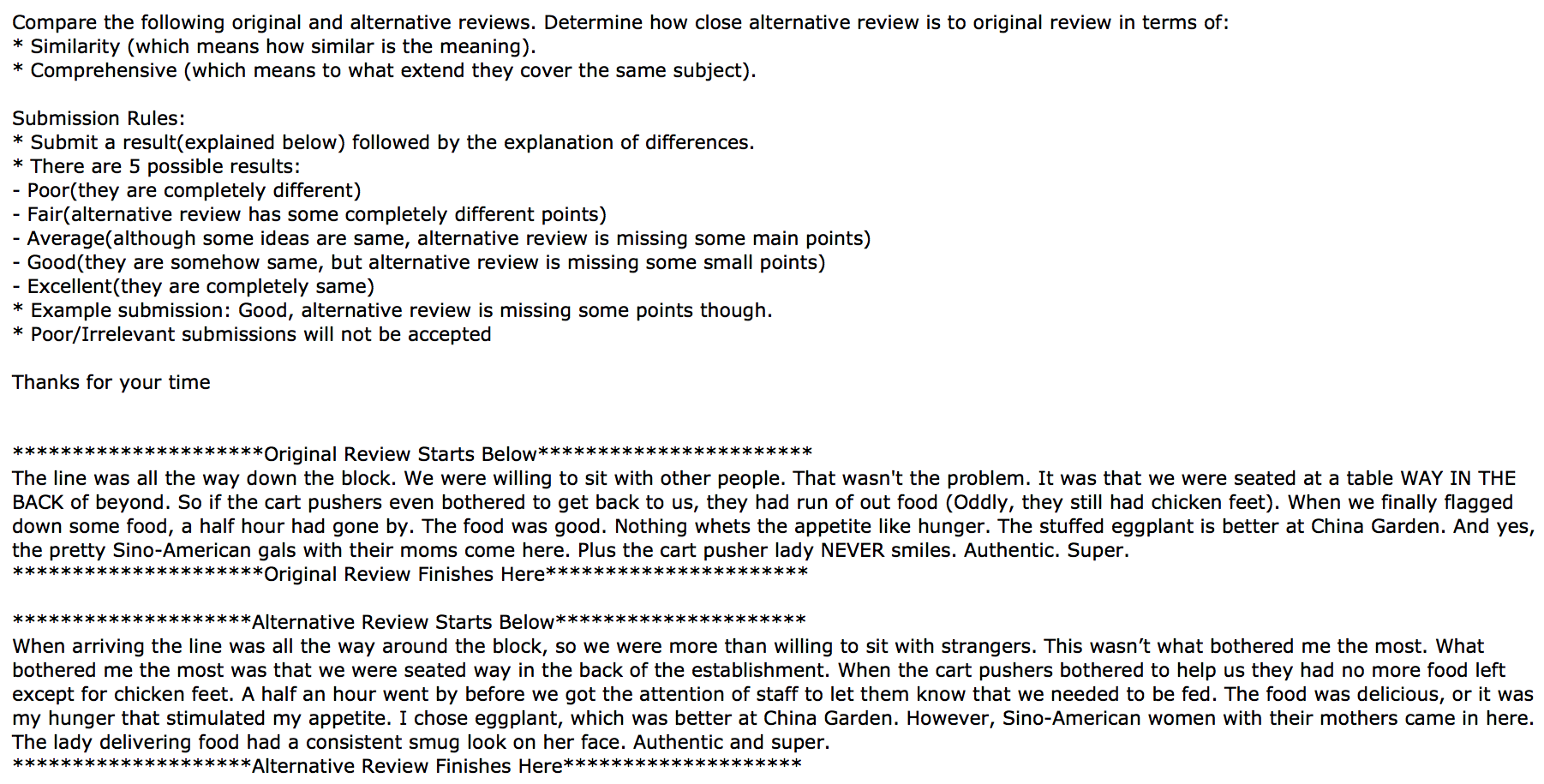}}}
  \caption{Examples Tasks in MTurk}
  \label{fig:example_tasks}
\end{figure*}

We present the round-trip machine translation of original review in
Figure \ref{fig:mt-rewrite-exp-png} with three, six and nine random 
selection of intermediate languages, and also translation-fixed version
by MTurk below:

\noindent{\textbf{Random round-trip machine translation with three
languages $English$ $\xrightarrow{}$ $Welsh$ $\xrightarrow{}$ $Irish$
$\xrightarrow{}$ $Ukrainian$ $\xrightarrow{}$ $English$: }\textit{
``The line was all the way down the block. We are ready to sit with other people.
This was not a problem. It was that we were sitting at the table, and the way in
the back longer. So if the cart pushers even bother to get back to us, they run
out of food (surprisingly, still chicken feet). When we finally gave up a little
food half an hour passed. Food was good. Nothing stimulates the appetite like
hunger. Best stuffed eggplants in the garden China. And many Chinese-American
girls and their mothers to come here. Plus a woman pushing a cart ever smiles.
Authentic. Super.''}}

\noindent{\textbf{Random round-trip machine translation with six
languages $English$ $\xrightarrow{}$ $Welsh$ $\xrightarrow{}$ $Irish$
$\xrightarrow{}$ $Ukrainian$ $\xrightarrow{}$ $Yiddish$ $\xrightarrow{}$
$French$ $\xrightarrow{}$ $Korean$ $\xrightarrow{}$ $English$:}\textit{
``The line at the bottom of the box is We sit down with other people who are
prepared. It was not a problem. Pusherz basket even if you do not bother coming
back to us, we have it sitting on the table, and how I was back in May, they
(surprisingly, many chicken feet) to food shortages. Some food when we passed
an hour and a half. The food was good. Nothing like hunger and appetite stimulation.
China stuffed in the garden of the. Many Chinese-American women and their mothers
come here. The woman pushing a cart does not smile. True. More.''}}

\noindent{\textbf{Random round-trip machine translation with nine
languages $English$ $\xrightarrow{}$ $Welsh$ $\xrightarrow{}$ $Irish$
$\xrightarrow{}$ $Ukrainian$ $\xrightarrow{}$ $Yiddish$ $\xrightarrow{}$
$French$ $\xrightarrow{}$ $Korean$ $\xrightarrow{}$ $Esperanto$ $\xrightarrow{}$
$Norwegian$ $\xrightarrow{}$ $Malay$ $\xrightarrow{}$ $English$: }\textit{
``Line at the bottom of the box, we sit down with other people who are willing.
It's not a problem. Pusherz cart even if you do not bother to come back to us,
we have it sitting on the table, and as I was back in May, (surprisingly, a lot
of chicken feet) to food shortages. Some foods when they pass an hour and a half.
Food is good. Nothing like hunger and appetite stimulation. China filling the park.
Many Chinese-American women and their mothers come here. Woman pushing a trolley
is not smiling. Sant. More.''}}

\noindent{\textbf{Fixed version of previous machine translated review with nine languages:}\textit{
``The problem wasn't that we weren't willing to sit with other people but the line
was still all the way down the block.  The problem was that we were seated a table
as far away as could be.  Even if the servers made it to our table, their trays were
empty (except for the chicken feet).  It was a half an hour before we were able to
get any of the food but it tasted good because we were so hungry; although, the China
Garden has better stuffed eggplant.  The pretty Sino-American girls come here with
their mothers and the woman server never smiles.  Traditional.  Awesome.''}}

Note that the full collection of translated reviews can be found in \cite{dropboxdataset}.

\begin{figure*}[t]
  \centering
  \fbox{\includegraphics[scale=0.45]{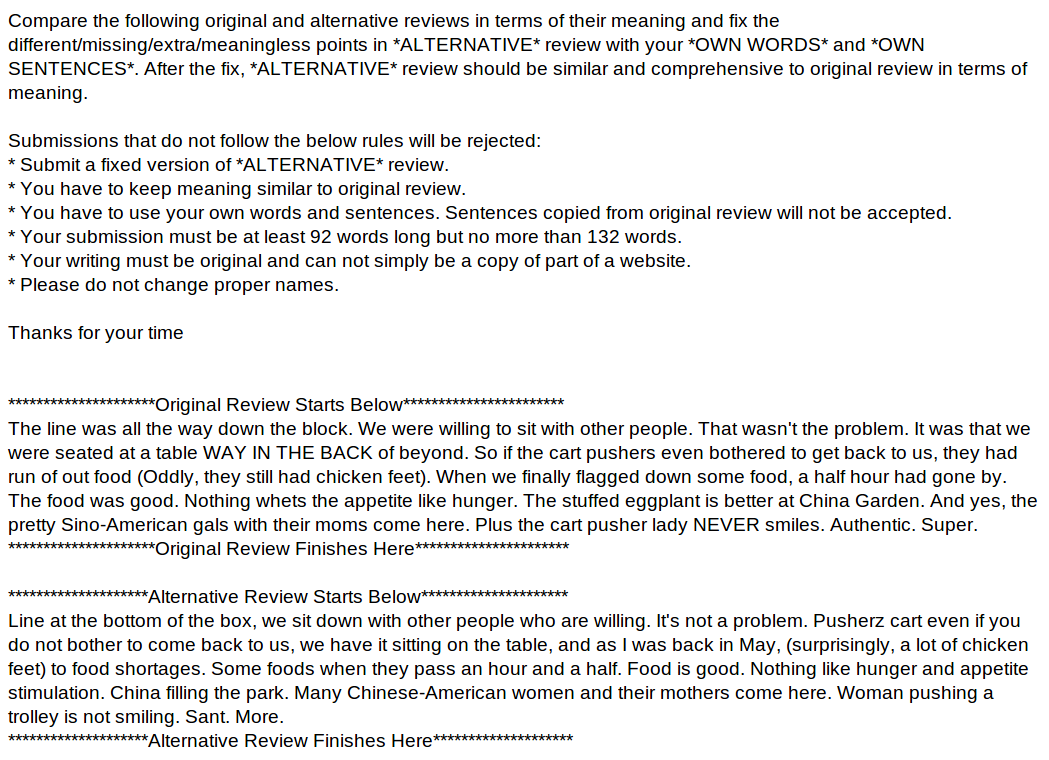}}
  \caption{Sample translation fix task in MTurk}
  \label{fig:example-translation-fix-box-png}
\end{figure*}